\documentclass[a4paper]{article}
\usepackage[utf8]{inputenc}
\usepackage[english]{babel}
\usepackage{authblk}
\usepackage{graphicx}
\usepackage{scrtime}
\usepackage{caption}
\usepackage[colorlinks]{hyperref}
\usepackage{etoolbox}
\usepackage[margin=3cm]{geometry}

\usepackage{amsmath,amssymb}
\usepackage{algorithm,algpseudocode}
\usepackage{dsfont}

\usepackage{parskip}

\begin{document}

\title{Casimir preserving numerical method for global multilayer geostrophic turbulence}

\author[1]{Arnout D. Franken}
\author[2]{Erwin Luesink}
\author[3]{Sagy R. Ephrati}
\author[1,4]{Bernard J. Geurts}
\affil[1]{Mathematics of Multiscale Modelling and Simulation, Department of Applied Mathematics, Faculty EEMCS, University of Twente, PO Box 217, 7500 AE Enschede, The Netherlands}
\affil[2]{Korteweg-de-Vries Institute, University of Amsterdam, PO Box 94248, Science Park 107, 1090 GE Amsterdam, The Netherlands}
\affil[3]{Department of Mathematical Sciences, Chalmers University of Technology, 412 96 Gothenburg, Sweden}
\affil[4]{Multiscale Physics, Center for Computational Energy Research, Department of Applied Physics, Eindhoven University of Technology, Eindhoven, The Netherlands}

\date{}

\maketitle

\begin{abstract}
Accurate long-term predictions of large-scale flow features on planets are crucial for understanding global atmospheric and oceanic systems, necessitating the development of numerical methods that can preserve essential physical structures over extended simulation periods without excessive computational costs. Recent advancements in the study of global single-layer barotropic models have led to novel numerical methods based on Lie-Poisson discretization that preserve energy, enstrophy and higher-order moments of potential vorticity. This work extends this approach to more complex stratified quasi-geostrophic (QG) systems on the sphere. 

In this work, we present a formulation of the multi-layer QG equations on the full globe. This allows for extending the Lie-Poisson discretization to multi-layer QG models, ensuring consistency with the underlying structure and enabling long-term simulations without additional regularization. The numerical method is benchmarked through simulations of forced geostrophic turbulence and the long-term behaviour of unforced multi-layered systems. These results demonstrate the structure-preserving properties and robustness of the proposed numerical method, paving the way for a better understanding of the role of high-order conserved quantities in large-scale geophysical flow dynamics.

\end{abstract}

\section{Introduction}

Long-time accurate predictions of large-scale flow features on planets are essential for understanding global climate systems~\cite{bechtold2016convection}. These predictions underpin a range of applications, from weather forecasting to the study of climate change, and play a significant role in informing policy and strategic decisions aimed at addressing climate impacts. 

A particular challenge in this field is the wide range of temporal and spatial scales involved in the full dynamic system. On the large planetary scales, an inverse energy cascade is present due to the two-dimensional nature of the flow at these scales, in which small eddies coalesce into increasingly larger eddies, thereby significantly contributing to the large-scale motion~\cite{chen2006physical}.

The limitations of existing models primarily stem from the need to parameterize submesoscale processes due to computational constraints. Traditional approaches often involve additional modelling efforts to approximate the effects of these unresolved scales, leading to increased complexity and potential inaccuracies~\cite{xiao2009physical,shevchenko2015multi}. This underscores the importance of developing numerical methods capable of preserving essential physical structures over long simulation times without necessarily relying on such approximations.

Recently, significant progress has been made through the development of structure-preserving numerical methods for single-layer barotropic models~\cite{cifani2023efficient, franken2024zeitlin}. These methods exploit the Lie-Poisson structure of the governing equations, allowing for discretizations that inherently maintain the essential structure of the equations such as conservation laws. As a result, these methods enable long-term simulations without the need for additional regularization, providing a self-contained framework for modelling. The efficacy of these structure-preserving methods has been demonstrated in accurately capturing the energy spectra of two-dimensional turbulence~\cite{cifani2022casimir}, as well as capturing the essential large-scale dynamics of quasi-geostrophic flows~\cite{franken2024zeitlin}.

Building on this foundation, the aim of this work is to extend the applicability of structure-preserving numerical methods to more complex, multi-layer quasi-geostrophic models. Multi-layered models are crucial for a more comprehensive representation of atmospheric and oceanic dynamics, as they account for vertical variations in density, pressure, and velocity fields~\cite{vallis2017atmospheric}. Extending these methods to multi-layer systems is a significant step towards capturing the intricate interactions between different layers of the atmosphere and ocean, as it extends the applicability of the models towards smaller scales.

In this paper, we present a detailed methodology for extending the structure-preserving Lie-Poisson discretization to multi-layer quasi-geostrophic models. Our approach leverages the same principles as in the single-layer models by ensuring that the discretization remains consistent with the underlying mathematical structures. We demonstrate that these methods can accurately simulate the long-term behaviour of multi-layer systems without requiring additional regularization.

By providing a benchmark for studying the influence of forcing and dissipative terms on large-scale dynamics, our work offers new insights into the role of submesoscale processes in shaping climate patterns. Additionally, we examine the impact of high-order moments of potential vorticity (PV) on large-scale systems, highlighting their significance for future developments in PV dynamics.

The structure of this paper is as follows. In Section~\ref{sec:Derivation}, we derive the multi-layer QG equations on the sphere from the equations of motion without relying on tangent plane approximations. In this section, we highlight the influence of spherical geometry on the governing equations and introduce the Lie-Poisson formulation of the equation. Using this formulation, we derive a numerical model for these equations in Section~\ref{sec:NumericalMethod} based on Zeitlin's method of quantisation. In Section~\ref{sec:Simulations}, we show numerical results from two test cases, showing the structure-preserving properties of the numerical method. Conclusions and an outlook are given in Section~\ref{sec:conclusion}.

\section{Derivation of the global multi-layer QG equations}\label{sec:Derivation}

The multi-layer QG equation describes the evolution of potential vorticity (PV). Starting from the Primitive Equations (PE), the QG equations are typically derived on a local rectangular tangent plane to the sphere by taking a variational expansion around the geostrophic balance (\cite{pedlosky2013geophysical}, \cite{vallis2017atmospheric}). On this tangent plane, the Coriolis parameter is approximated as a constant or as a linear function of the latitudinal coordinate. This leads to the so-called $f$-plane and $\beta$-plane approximations respectively. The use of such a local domain implies the introduction of nonphysical boundary conditions. In this work, we will consider a global QG model, defined directly on the entire sphere. This enables the inclusion of the fully latitude-dependent Coriolis parameter, as well as the spherical geometry of the domain.

Previously, it was shown independently by \cite{schubert2009shallow} and \cite{verkley2009balanced} that a global single-layer QG model can be derived from shallow water models by a solenoidal partitioning of the flow that distinguishes between a divergent and a non-divergent part. This provides an alternative to the traditional partitioning into a geostrophic and an ageostrophic component, which is ill-defined at the equator due to the vanishing Coriolis force. In this section, we follow the solenoidal partitioning of the flow and derive the associated multi-layer QG model on the sphere. This multi-layer model makes it possible to incorporate many essential phenomena that are not included in the single-layer QG model. Crucially, the single-layer QG model treats the fluid as a single homogeneous layer on top of a stationary surface. By including vertical variations in buoyancy, the multi-layer model includes internal baroclinic modes, which allows for the study of baroclinic instabilities~\cite{phillips1963geostrophic, salmon1980baroclinic}.

The derivation of the multi-layer QG equations starts from the Boussinesq primitive equation (PE) (Subsection~\ref{subsec:PE}), to which an approximate quasi-geostrophic balance is imposed in case of stable stratification (Subsection~\ref{subsec:StratifiedQG}), from which a multi-layer model can be derived~(Subsection~\ref{subsec:multi-layer}).

\subsection{Boussinesq primitive equations on the sphere}\label{subsec:PE}

The stratified QG equations can be derived from the Boussinesq PE, which describes the motion of a stratified hydrostatic fluid layer. A comprehensive derivation from Newton's equation of motion is provided in \cite{pedlosky2013geophysical} and \cite{vallis2017atmospheric} on a tangent plane of the sphere. In this section, we use this approach to similarly derive the Boussinesq PE on a spherical domain from which the stratified QG equations can then be derived.

We consider a shallow layer of incompressible fluid on a sphere of radius $R$ rotating with angular velocity $\mathbf{\Omega}$ obeying Newton's equation of motion. The assumption of a perfect sphere upon which a fluid layer is deposited that is very shallow compared to the radius of the sphere, yields by good approximation a uniform and constant gravitational acceleration $-g$ in the radial direction~\cite{zeitlin2018geophysical}. Thus, we may write the equations of motion as
\begin{align}
    &\frac{\partial \mathbf{v}}{\partial t} + \mathbf{v }\cdot \nabla \mathbf{v} + 2 \mathbf{\Omega} \times \mathbf{v} + g \mathbf{\hat{r}} = -\frac{1}{\rho}\nabla p, \label{eq:NewtonEquations}\\
    &\frac{\partial \rho}{\partial t} + \nabla \cdot (\rho \mathbf{v} ) = 0,
   \label{eq:densityequation}
\end{align}
where $\mathbf{v}$ is the velocity field, $p$ is the pressure and $\rho$ the mass density, which are functions of time $t$ and space ${\mathbf{x}}$. Since the fluid domain is a spherical shell, we use spherical coordinates to denote position as $\mathbf{x}=(r,\phi,\theta)$, where $r$ is the radial coordinate, and $\phi$ and $\theta$ denote latitude and longitude respectively. Moreover, gravitational acceleration is directed in the radial direction with unit vector $\mathbf{\hat{r}}$.

In the sequel, these equations are subjected to three basic assumptions that are adopted with respect to large-scale planetary flow:
\begin{enumerate}
    \item the centrifugal force due to rotation is small compared to the gravitational acceleration;
    \item in planetary flows, the vertical velocity component is small compared to horizontal velocities;
    \item inertial accelerations are small compared to gravitational acceleration.
\end{enumerate}
These assumptions follow from scale comparisons in common planetary flows~\cite{pedlosky2013geophysical} and will be quantified in the following discussions. As a result of these assumptions, large-scale motion in the radial momentum equation is close to hydrostatic equilibrium, which implies
\begin{equation}
    \frac{\partial p}{\partial r} = -\rho g.
    \label{eq:hydrobalance}
\end{equation}
This provides the governing equation from which the pressure can be determined. Furthermore, we assume that the density field is close to stable stratification with a radial density profile of $\overline{\rho}(r)$ around a reference value $\rho_0$. Following the approach by \cite{zeitlin2018geophysical}, we thus decompose the density field in three parts
\begin{equation}
    \rho(\mathbf{x},t) = \rho_0 + \overline{\rho}(r) + \rho'(\mathbf{x},t),
    \label{eq:densitydecomposition}
\end{equation}
where $\overline{\rho}$ gives the average stratification and $\rho'$ denotes small density fluctuations. The assumption of approximately stable stratification then implies that $\overline{\rho}(r)$ decreases monotonically with increasing $r$ and that the density fluctuations must be small compared to the stratification, leading to the relations $||\rho'||\ll ||\overline{\rho}|| \ll \rho_0$. From equation (\ref{eq:hydrobalance}), it now follows that the pressure can similarly be decomposed into three parts as
\begin{equation}
    p = \rho_0 g r + \overline{p}(r) + p'(\mathbf{x},t),
\end{equation}
in which each term on the right-hand side corresponds to the respective term in the density decomposition. Furthermore, substituting the density decomposition (\ref{eq:densitydecomposition}) into the continuity equation (\ref{eq:densityequation}) yields

\begin{equation}
    \frac{\partial \rho'}{\partial t} + w\frac{\partial \overline{\rho}}{\partial r} + \mathbf{v} \cdot \nabla \rho' + \rho(\nabla \cdot \mathbf{v}) = 0
    \label{eq:continuity_decomposed}
\end{equation}

This implies that the divergence of the velocity field is very small, specifically, of the order $|\rho'|/\rho_0$ \cite{zeitlin2018geophysical}. Thus, we additionally require that the total velocity field is incompressible, which amounts to the classical Boussinesq approximation. Neglecting the small density variations in the right-hand side of (\ref{eq:NewtonEquations}) now gives the primitive equations (PE) in the Boussinesq approximation on the sphere: 
\begin{align}
    \frac{\partial \mathbf{v}_h}{\partial t} + \mathbf{v} \cdot \nabla \mathbf{v}_h + f \hat{\mathbf{r}} \times \mathbf{v}_h &= - \frac{1}{\rho_0}\nabla_h p',\label{eq:horMomentumPE}\\
    \frac{\partial p'}{\partial r} + \rho' g &= 0, \label{eq:hydrostaticPE}\\
    \frac{\partial \rho'}{\partial t} + \mathbf{v}\cdot \nabla \rho' + w\frac{\partial \overline{\rho}}{\partial r}&= 0, \label{eq:thermoinPE}\\
    \nabla_h \cdot \mathbf{v}_h + \frac{1}{r^2}\frac{\partial r^2w}{\partial r} &= 0,
\end{align}
where $f = 2\Omega \cos{\phi}$ is the Coriolis parameter with $\Omega$ the magnitude of the angular velocity vector. The first two equations follow from the tangential (horizontal) and radial momentum equations respectively, while the third equation follows from the mass transport (\ref{eq:continuity_decomposed}) using the incompressibility of the velocity field. This latter condition is included in the fourth equation. Together, they form the familiar Boussinesq PE on the sphere~\cite{vallis2019essentials}. The velocity vector $\mathbf{v}$ is split into its radial component $w$ and the horizontal velocity field $\mathbf{v}_h$, given by $\mathbf{v} = \mathbf{v}_h + w \hat{\mathbf{r}}$, such that $\mathbf{v}_h\cdot \mathbf{\hat{r}} = 0$. To highlight this decomposition further, we write the del operator as $\nabla_h$ in case all radial derivatives are zero.  Typically, the average density stratification is given which implies that these equations provide a closed system in terms of the velocity field and the fluctuating parts of density and pressure. In the next subsection, we use scaling arguments to derive the stratified QG equations on the sphere.

\subsection{Stratified quasi-geostrophic equations on the sphere}\label{subsec:StratifiedQG}

The PE as formulated above constitute a complex model of the dynamics of a shallow fluid layer on a rapidly rotating planet. In the study of large-scale motions of this fluid, it appears that a large separation exists in the magnitude of the independent variables in the equations. To examine this, we consider scaling these equations and selecting a length scale for the horizontal motion $L$, as well as a horizontal velocity scale $U$. The length scale for vertical motion is the mean height of the fluid column, denoted by $H$, for which a vertical velocity scale $W$ can be determined as $W=UH/L$. Since we are mainly interested in the horizontal motion, we introduce a time scale $L/U$. For the rotation of the sphere, the scale is given by $\Omega$, while the density scale has already been given as $\rho_0$. We now assume that the parameters are such that the so-called global Rossby number is small, i.e., $\mbox{Ro}=U/L\Omega\ll 1$, which holds for sufficiently large-scale horizontal motion. This is the core assumption for geostrophic flow, which implies that inertial forces are small compared to the Coriolis and gravitational forces \cite{phillips1963geostrophic}. This yields a balance between the latter two forces, such that the momentum equation (\ref{eq:horMomentumPE}) yields
\begin{equation}
    \left|f\hat{\mathbf{r}}\times \mathbf{v}_h\right| \sim \left|\frac{1}{\rho_0} \nabla_h p'\right|
\end{equation}
It is important to note that this does not hold in a small equatorial region on the sphere where $\hat{\mathbf{r}}\times \mathbf{v}_h$ goes to zero. In other words, the motion in that region is not strictly geostrophic. Away from the equator, this balance is however quite accurate, which leads to a natural scaling of $p'\sim\rho_0\Omega U L$. The hydrostatic balance (\ref{eq:hydrostaticPE}) then yields the scaling of density fluctuations $\rho'\sim\rho_0\Omega UL/gH$.

Finally, as is customary for stratified Boussinesq flows, we rewrite the average vertical density stratification by the Brunt-Väisälä frequency $N$, defined as
\begin{equation}\label{eq:bruntvaisala}
    N(r) =\sqrt{ -\frac{g}{\rho_0}\frac{\partial \overline{\rho}}{\partial r}},
\end{equation}
which represents the oscillating frequency of a displaced fluid parcel in a density gradient. This allows us to write the continuity equation (\ref{eq:thermoinPE}) as
\begin{equation}
    \frac{\partial \rho'}{\partial t} + \mathbf{v} \cdot \nabla \rho' - \frac{\rho_0}{g}wN^2 = 0.
\end{equation}
With the above scaling, including a typical Brunt-Väisälä frequency $N_0$ at mid-latitudes, we can non-dimensionalise the PE in terms of non-dimensionalised variables denoted with a $\thicksim$ as
\begin{align}
    \mbox{Ro}\left(\frac{\partial \mathbf{\Tilde{v}}_h}{\partial \Tilde{t}} + \mathbf{\Tilde{v}} \cdot \Tilde{\nabla} \mathbf{\Tilde{v}}_h\right) + \Tilde{f} \hat{\mathbf{r}} \times \mathbf{\Tilde{v}}_h &= - \Tilde{\nabla}_h \Tilde{p}',\label{eq:horMomentumPEnondim}\\
    \frac{\partial \Tilde{p}'}{\partial \Tilde{r}} + \Tilde{\rho}' &= 0, \label{eq:hydrostaticPEnondim}\\
     \mbox{Ro}\left( \frac{\partial \Tilde{\rho}'}{\partial \Tilde{t}} + \mathbf{\Tilde{v}}\cdot \Tilde{\nabla} \Tilde{\rho}' \right) + \left(\frac{L_d}{L}\right)^2 \Tilde{w}\Tilde{N}^2 &= 0, \label{eq:thermoinPEnondim}\\
    \Tilde{\nabla}_h \cdot \mathbf{\Tilde{v}}_h + \frac{\partial \Tilde{w}}{\partial \Tilde{r}} &= 0.\label{eq:divergencefreeinPEnondim}
\end{align}
Furthermore, we have introduced the baroclinic Rossby deformation radius $L_d$, given by
\begin{equation}
    L_d = \frac{N_0H}{\Omega},
\end{equation}
which is analogous to the deformation radius $\sqrt{gH}/\Omega$ in shallow water systems, such as that seen in \cite{verkley2009balanced} and \cite{franken2024zeitlin}. 

One important aspect to highlight is that, although the equations are valid everywhere, this particular scaling is not strictly applicable in a narrow band around the equator. At very small latitudes, the Coriolis term decreases, and the pressure scaling can no longer be determined from equation (\ref{eq:horMomentumPEnondim}) as $\Tilde{f}$ becomes of the same order as the Rossby number. Specifically, the ratio between inertial forces and the Coriolis force is locally given by the local Rossby number $\mbox{Ro}_{loc} = U/L\Omega \sin \phi = \mbox{Ro}/\sin\phi$. In the regime of rapid rotation that we consider, the global Rossby number is very small, leading to a similarly small local Rossby number at moderate and high latitudes. The domain where the scaling is not applicable is thus contained in a very narrow equatorial band where the local Rossby number requirement for geostrophic flow $\mbox{Ro}_{loc}\ll 1$ is not satisfied. In the following, we will continue with the non-dimensionalised equations, but the tildes will be omitted for clarity.

\subsection{Geostrophic balance}

In the regime of rapid rotation where the Rossby number is small, it can be seen directly from the momentum equation (\ref{eq:horMomentumPEnondim}) that inertial forces are small, and that there is an approximate balance between the Coriolis force and the hydrostatic pressure force. In this case of strong scale separation, the flow is said to be in geostrophic balance \cite{vallis2019essentials}. In this section, we discuss the impact of the scale separation on the dynamics of the Boussinesq PE.

The non-dimensionalized Boussinesq PE reveal that the dynamics depend on the considered horizontal length scale $L$. The quasi-geostrophic approximation can be applied when the considered horizontal length scale of the flow is of the same order as the Rossby deformation radius, or, $L_d/L = \mathcal{O}(1)$. Typical values of $L_d$ in geophysical flows depend on the application area. For example, in Earth's atmosphere this scale is typically of $\mathcal{O}(10^3)$ km, while in the Earth's upper ocean, this is of $\mathcal{O}(10-10^2)$ km\cite{vallis2019essentials}. Another possible application area is the atmosphere of Jupiter, for which $L_d\sim 2000$ km \cite{luesink2024geometric}. It should be noted that, especially in the latter two cases, we can consider length scales for which $L_d/L=\mathcal{O}(1)$ while having $L/R\ll 1$. 

In a tangent plane approximation, the geostrophic balance is used to decompose the horizontal velocity field into its geostrophic and ageostrophic parts. The geostrophic velocity is found using the scale separation equation (\ref{eq:horMomentumPEnondim}) by neglecting the inertial term:
\begin{equation}\label{eq:geostrophicbalance}
    f \hat{\mathbf{r}}\times \mathbf{v}_h + \nabla_h p' = 0.
\end{equation}
However, this geostrophic velocity is ill-defined on the equator~\cite{vallis2017atmospheric}. On the sphere, we therefore do not explicitly calculate this geostrophic velocity. Taking the cross-derivative of the balance equation above, defined as a left-multiplication by $\hat{\mathbf{r}}\cdot\nabla \times$ gives
\begin{equation}\label{eq:geostrophicdivergence}
    f \nabla_h\cdot \mathbf{v}_h + \mathbf{v}_h \cdot \nabla_h f = 0.
\end{equation}
where we use the fact that the curl of a gradient field is zero everywhere. The first term contains the divergence of the horizontal part of the velocity field, while the second term contains the gradient of the nondimensional Coriolis parameter. Due to the non-dimensionalization, this is $\mathcal{O}(L/R)$. Thus, for the horizontal length scales around the baroclinic Rossby radius, this is a small quantity. The balance above thus implies that the horizontal velocity field is divergence-free to a good approximation. Instead of the typical decomposition of the velocity field into a geostrophic and an ageostrophic component, here we split the field into a dominant divergence-free field and a remaining term of small magnitude. Using the Helmholtz decomposition, we can therefore introduce a streamfunction $\psi$ for the former part and write the velocity decomposition as
\begin{equation}\label{eq:velocitydecomposition}
    \mathbf{v}_h = \hat{\mathbf{r}}\times \nabla \psi + \epsilon \mathbf{v}_{h}',
\end{equation}
where $\epsilon \ll 1$ is a scaling parameter for the divergent part of the horizontal velocity field $\mathbf{v}_h'$. Since the total velocity field is still assumed to be divergence-free, we find from (\ref{eq:divergencefreeinPEnondim}) that 
\begin{equation}\label{eq:verticalvelocityscaling}
    \frac{\partial w}{\partial r} + \epsilon \nabla_h \cdot \mathbf{v}_h' = 0.
\end{equation}
This implies that the vertical velocity is smaller than the suggested scaling of $UH/L$ due to the presence of the small parameter $\epsilon$. From this point, we can follow the same steps as in typical derivations on a local tangent plane. By cross-differentiating the horizontal momentum equation, i.e., applying $\mathbf{\hat{r}}\cdot \nabla \times $ to equation (\ref{eq:horMomentumPEnondim}), we find
\begin{equation}\label{eq:crossdiffMomentum}
    \mbox{Ro} \left[ \frac{\partial \zeta}{\partial t} + \mathbf{v}_h \cdot \nabla_h \zeta +\zeta \nabla_h \cdot \mathbf{v}_h +  w \frac{\partial \zeta}{\partial r} - \mathbf{\hat{r}} \cdot \nabla_h w \times \frac{\partial \mathbf{v}_h}{\partial r} \right]+ \mathbf{v}_h\cdot \nabla_h f + f \nabla_h\cdot\mathbf{v}_h = 0,
\end{equation}
where we introduced the (scalar-valued) vorticity of the horizontal velocity field $\zeta := \hat{\mathbf{r}}\cdot \nabla \times \mathbf{v}_h$. Given the scaling assumptions above, we can neglect the fourth and fifth terms in the square brackets, since the vertical velocity is of $\mathcal{O}(\epsilon)$. Similarly, the third term is neglected due to the small divergence of the horizontal velocity. In the remaining terms, we now substitute the velocity decomposition (\ref{eq:velocitydecomposition}) and neglect the non-divergent part in the advective terms. Defining a horizontal material derivative $\mbox{D}/\mbox{D}t(\cdot) = \partial(\cdot)/\partial t + \hat{\mathbf{r}}\times \nabla \psi \cdot \nabla_h (\cdot)$ now gives
\begin{equation}
    \frac{\mbox{D}}{\mbox{D}t}(\mbox{Ro}\,\zeta + f) + \epsilon (\mbox{Ro}\,\zeta + f)\nabla_h \cdot \mathbf{v}_h' = 0, \label{eq:temp}
\end{equation}
using the fact that the Coriolis parameter is time-independent. Since the Rossby number is small, we can neglect the first term in the second brackets as it includes the small parameter $\epsilon$, and we introduce the divergence-free requirement (\ref{eq:verticalvelocityscaling}) to find
\begin{equation}
    \frac{\mbox{D}}{\mbox{D}t}(\mbox{Ro}\,\nabla^2\psi + f) -  f \frac{\partial w}{\partial r}= 0,\label{eq:temp2}
\end{equation}
where we use the small value of $\epsilon$ in equation (\ref{eq:velocitydecomposition}) to approximate the horizontal vorticity by the Laplacian of the streamfunction $\psi$. We can use the continuity equation (\ref{eq:thermoinPEnondim}) to eliminate the vertical velocity. By again neglecting the nondivergent part of the advecting horizontal velocity, we can rewrite the continuity equation as
\begin{equation}
    \mbox{Ro}\frac{\mbox{D}}{\mbox{D}t}\rho' + \left(\frac{L_d}{L}\right)^2 wN^2 = 0.
\end{equation}
The above equation can be used to determine the vertical velocity $w$ to be used in the vorticity evolution (\ref{eq:temp}). By noting that the radial derivative commutes with the horizontal material derivative, we can extract the vertical velocity by taking the radial derivative to find

\begin{equation}
\begin{split}
    f\frac{\partial w}{\partial r} &= -\mbox{Ro}\left(\frac{L}{L_d}\right)^2 f\frac{\mbox{D}}{\mbox{D}t}\left(\frac{\partial}{\partial z}\left(\frac{\rho'}{N^2}\right) \right) \\
    &= -\mbox{Ro}\left(\frac{L}{L_d}\right)^2 \left[ \frac{\mbox{D}}{\mbox{D}t}\left(f\frac{\partial}{\partial z}\left(\frac{\rho'}{N^2}\right) \right)- \frac{\partial}{\partial z}\left(\frac{\rho'}{N^2}\right)\frac{\mbox{D}f}{\mbox{D}t}\right]
\end{split}
\end{equation}

The second term between square brackets can be neglected since the Coriolis term $f$ varies only slowly over the sphere which implies $\mbox{D}f/\mbox{D}t \sim \mathcal{O}(L/R)$. Combining equations (\ref{eq:geostrophicbalance}) and (\ref{eq:velocitydecomposition}), we find that the pressure fluctuations are predominantly determined by the divergence-free part of the horizontal velocity field, such that we can approximate $p' = f \psi$ (up to a gauge constant). This result, combined with the hydrostatic pressure relation (\ref{eq:hydrostaticPEnondim}) allows us to write the density fluctuations in terms of the streamfunction, resulting in
\begin{equation}
    f\frac{\partial w}{\partial r} = -\mbox{Ro}\left(\frac{L}{L_d}\right)^2 \frac{\mbox{D}}{\mbox{D}t}\left( f^2 \frac{\partial}{\partial r} \left( \frac{1}{N^2}\frac{\partial \psi}{\partial r}\right)\right),
\end{equation}
where we use the fact that the Coriolis parameter is independent of the radial coordinate in the shallow water assumption. Substituting this result in the dynamical equation (\ref{eq:temp2}) now gives stratified QG equations on the sphere
\begin{equation}\label{eq:stratifiedQGfinal}
    \frac{\mbox{D}q}{\mbox{D}t} = 0, \quad q = \mbox{Ro}\,\nabla^2\psi + f + \mbox{Ro}\left(\frac{L}{L_d}\right)^2 f^2 \frac{\partial}{\partial r} \left( \frac{1}{N^2}\frac{\partial \psi}{\partial r}\right),
\end{equation}
where $q$ is the potential vorticity for the stratified QG equations, which is materially conserved on horizontal planes and is the single prognostic variable. This final result is very similar to the traditional stratified QG equation on a Cartesian plane, except for the crucial appearance of the fully latitude-dependent Coriolis term $f$. The main difference in the derivation lies in the decomposition of the horizontal velocity in terms of a divergence-free contribution and a small divergent field. The geostrophic velocity can now be reconstructed from the streamfunction, which is well-defined on the entire sphere.

Even though the final formulation of the stratified QG equations on the sphere appears to be a straightforward generalization of the well-known formulation on the $\beta$-plane, its derivation sheds some light on the applicability of the model. Starting from the Boussinesq PE on the sphere, the geostrophic balance is used to derive the leading order dynamics in terms of the Rossby number. This is valid at all latitudes where a scale separation exists between the Coriolis force and inertial forces. It thus provides a continuous model for both mid-latitude and polar dynamics. At a narrow band around the equator, scale separation breaks down as the Coriolis parameter vanishes. In the absence of a local geostrophic balance, gravity waves play an important role and the QG equation does not provide accurate results for the dynamics~\cite{vallis2017atmospheric}.

\subsection{Multi-layer stratification}\label{subsec:multi-layer}

From the continuous equations, a layered model may be derived by subdividing the domain horizontally. Within each layer, a constant density may be set, such that the density distribution over the layers approximates the continuous stratification that would appear in a full 3D formulation. This is equivalent to employing a finite-difference discretization in the radial direction~\cite{vallis2019essentials}, which will be the approach in this section. When the average density stratification is stable, its level sets (isopycnals, or surfaces of constant density) form non-overlapping horizontal slices of the fluid column. In the derivations above, we assumed that density fluctuations are small compared to the average stratification, indicating that the isopycnals form approximately form spherical shells. Thus we describe the fluid using several layers of constant thickness and uniform density given by the average stratification, whose interfaces correspond to isopycnals of the continuous model. 

Using this approach, we divide the domain in the radial direction into $M$ spherical shells. We denote the layers from top to bottom with index $j=1,\ldots,M$, where each layer has thickness $H_j$ such that $\sum_{j=1}^M H_j = H$. We can describe the dynamics in each layer by the potential vorticity $q_j$ and streamfunction $\psi_j$. 

We employ a second-order centred finite difference scheme to discretize the radial derivatives in (\ref{eq:stratifiedQGfinal}), similar to the approach taken in \cite{vallis2019essentials} for the $\beta$-plane equations. This yields the approximation
\begin{equation}\label{eq:finitediff}
    \frac{\partial}{\partial r}\left( \frac{1}{N^2} \frac{\partial \psi}{\partial r} \right) \approx \frac{1}{H_j}\left( \frac{1}{N_{j-\frac{1}{2}}^2}\frac{\psi_{j-1} - \psi_j}{H_{j-\frac{1}{2}}} - \frac{1}{N_{j+\frac{1}{2}}^2}\frac{\psi_{j} - \psi_{j+1}}{H_{j+\frac{1}{2}}}   \right),
\end{equation}
where $H_{j+\frac{1}{2}} = (H_{j+1}+H_j)/2$ and the Brunt-Väisälä frequencies are given by central differencing of equation (\ref{eq:bruntvaisala}) as
\begin{equation}\label{eq:bruntvaisaladiff}
    N_{j+\frac{1}{2}}^2 = - \frac{g}{H_{j+\frac{1}{2}}}\frac{\rho_{j}-\rho_{j+1}}{\rho_0} = - \frac{g_{j+\frac{1}{2}}'}{H_{j+\frac{1}{2}}}, \quad \mbox{where} \quad g'_{j+\frac{1}{2}} = g\frac{\rho_{j+1}-\rho_j}{\rho_0}
\end{equation}
where the reduced gravity $g_{j+\frac{1}{2}}'$ has been introduced as a measure of the effective gravitational acceleration between layers. Applying this discretization in the radial direction to the stratified QG equations (\ref{eq:stratifiedQGfinal}), we find the multi-layer QG equations on the sphere
\begin{equation}\label{eq:PVtransportPoisson}
    \Dot{q}_j + \{ \psi_j, q_j\} = 0, \quad j=1,\ldots, M
\end{equation}
where $q_j$ is the layerwise potential vorticity, given by 
\begin{equation} \label{eq:multi-layerPsi}
    \mathbf{q} = \nabla^2 \boldsymbol{\psi} + \mathbf{f} + A \boldsymbol{\psi},
\end{equation}
where $\mathbf{q} = [q_1,q_2,\ldots,q_M]^\top$ is a vector containing the layer-wise PV, and similarly $\boldsymbol{\psi} = [\psi_1,\psi_2,\ldots,\psi_M]^\top$ and $\mathbf{f} = [1,1,\ldots, 1]^\top f$. The interaction between layers is given by the matrix $A\in \mathds{R}^{M\times M}$, which is given in terms of the reduced gravity and layer thickness as
\begin{equation}\label{eq:interactionmatrix}
    A = \mbox{Ro}\left(\frac{L}{L_d}\right)^2 f^2 \begin{bmatrix}
        -\frac{1}{g_{3/2}'H_1} & \frac{1}{g_{3/2}'H_1} & 0 & \ddots & 0 \\
        \frac{1}{g_{3/2}'H_2} & -\frac{1}{g_{3/2}'H_2}- \frac{1}{g_{5/2}'H_2} & \frac{1}{g_{5/2}'H_2} & \ddots & 0\\
        0 & \frac{1}{g_{5/2}'H_3} & -\frac{1}{g_{5/2}'H_4}- \frac{1}{g_{7/2}'H_4} & \ddots & 0 \\
        \vdots & \ddots & \ddots & \ddots & \vdots \\
        0 & \cdots & \cdots & \cdots & -\frac{1}{g_{M-1/2}'H_L}
    \end{bmatrix}
\end{equation}
as follows directly from (\ref{eq:stratifiedQGfinal}) using (\ref{eq:finitediff}) and (\ref{eq:bruntvaisaladiff}). The horizontal material derivative $\mbox{D}/\mbox{D}t$ has been rewritten using the canonical Poisson bracket on the sphere, which, in terms of the latitude $\phi$ and the longitude $\theta$ is given by
\begin{equation}\label{eq:PoissonBracketDefinition}
    \{a, b \} = \frac{1}{R^2\cos\phi}\left( \frac{\partial a}{\partial \theta} \frac{\partial b}{\partial \phi} - \frac{\partial a}{\partial \phi}\frac{\partial b}{\partial \theta}\right) = \left(\hat{\mathbf{r}}\times \nabla a\right) \cdot \nabla b.
\end{equation}

From this formulation of horizontal PV transport in terms of the canonical Poisson bracket on the sphere, we note the layerwise similarity to the single-layer QG equation as described on the sphere \cite{verkley2009balanced} and \cite{schubert2009shallow}. In particular, the only difference appears in the relation between streamfunction and potential vorticity (\ref{eq:multi-layerPsi}). 

Recently, the Poisson structure of the single-layer QG equation on the sphere has been discussed in \cite{franken2024zeitlin}. From their analysis, we find that (\ref{eq:PVtransportPoisson}) constitutes a Lie-Poisson system on the space of smooth functionals on the sphere $\mathcal{C^\infty}(\mathcal{S}^2)$ for each layer separately, where the Lie-Poisson bracket is given in terms of the canonical Poisson bracket by 
\begin{equation}\label{eq:LiePoissonBracketInfdim}
    \{ \mathcal{A}, \mathcal{B} \}_{LP}(q) = \int_{\mathcal{S}^2} q \left\{ \frac{\delta \mathcal{A}}{\delta q}(q), \frac{\delta \mathcal{B}}{\delta q}(q) \right\} \mbox{d}x,
\end{equation}
with $\mathcal{A},\mathcal{B}:\, \mathcal{C}^\infty(\mathcal{S}^2)\to \mathds{R}$ smooth functionals and $\mbox{d}x$ is an infinitesimal area element, which in latitude-longitude coordinates is given by $\mbox{d}x=R\cos\phi \,\mbox{d}\phi \,\mbox{d}\theta$. The advection of layer-wise PV (\ref{eq:PVtransportPoisson}) may thus be more generally written as
\begin{equation}\label{eq:LiePoissonContinuous}
    \frac{\mbox{d}}{\mbox{d}t}\mathcal{F}(q_j) = \{ \mathcal{F}(q_j), \mathcal{H}(q_j) \}_{LP},
\end{equation}
with the Hamiltonian functional given by
\begin{equation}\label{eq:HamiltonianContinuous}
    \mathcal{H}(q) = -\frac{1}{2}\sum_{j=1}^M \int_{\mathcal{S}^2} \psi_j (q_j-f) \, \mbox{d}x
\end{equation}
Thus, similar to the single-layer QG equations, this system allows for infinitely many conserved quantities in the form of Casimir functionals $\mathcal{C}:C^\infty(\mathcal{S}^2)\to\mathds{R}$ for which holds
\begin{equation}\label{eq:CasimirDefinition}
    \{\mathcal{F}(q_j),\mathcal{C}(q_j)\}_{LP} = 0, \quad \forall \mathcal{F}: C^\infty(\mathcal{S}^2)\to\mathds{R},
\end{equation}
which in this case results in $\mathcal{C}(q_j)=\int_{S^2} \zeta(q_j)$ for any smooth function $\zeta\in C^\infty(\mathds{R})$, which indicates that for each layer, any integrated function of potential vorticity is a constant of motion. This may be verified by direct computation
\begin{equation}\label{eq:CasimirConservation}
\begin{split}
    \frac{\mbox{d}}{\mbox{d}t}\mathcal{C}(q_j) &= \int_{\mathcal{S}^2} \dot{q_j} \zeta'(q_j)\mbox{d}x = - \int_{\mathcal{S}^2} \left((\mathbf{u}_j)_h \cdot \nabla q_j\right) \zeta'(q_j)\mbox{d}x \\
    &= - \int_{\mathcal{S}^2} (\mathbf{u}_j)_h \cdot \nabla \zeta(q_j)\mbox{d}x = \int_{\mathcal{S}^2}\nabla\cdot(\mathbf{u}_j)_h \zeta(q_j)\mbox{d}x \\
    &= 0.
\end{split}
\end{equation}
Typically, only integrated monomials of PV are considered since they form a basis for functions on the sphere. In particular, the integrated square of PV is called the total enstrophy and plays an important role in the characterization of geophysical fluid dynamics~\cite{salmon1978twolayer,gkioulekas2012effect}. As previously noted in \cite{franken2024zeitlin}, this conservative property is independent of the specific relation between streamfunction and potential vorticity. This explains the similarity in the constants of motion between this model and other incompressible 2D flows such as the 2D Euler equations and the rotating shallow water equations. For a detailed overview of the relation between these models, we refer to \cite{luesink2021stochastic}.




In the next section, we use the approach of \cite{zeitlin2004self} to apply a spatial discretization that preserves as many Casimirs of the continuous system as possible. By choosing a suitable time integrator, these Casimirs may then be conserved in a numerical solution. This allows us to numerically integrate the full multilayer system (\ref{eq:multi-layerPsi}) whilst preserving the underlying mathematical structure, thereby enabling the study of, e.g., the role of higher-order moments of PV on the flow evolution.

\section{Numerical method}\label{sec:NumericalMethod}

In the previous section, the multi-layered QG equations on the sphere were derived, resulting in a dynamical equation for potential vorticity $q$, in terms of the streamfunction $\psi$. This system features an infinite family of constants of motion due to the Lie-Poisson property of the dynamical system. In this section, we construct a numerical method for solving this system that aims to preserve the Lie-Poisson structure in order to retain as many constants of motion as possible in the numerical solution. This approach comes in two steps. First, we project functions on the sphere onto a particular discrete basis such that the resulting system of ODEs is Lie-Poisson. Then, a suitable time integrator is selected to conserve the Casimirs of the discrete system as well as the associated Hamiltonian energy.

\subsection{Spatial discretization}

The construction of a projection of the infinite-dimensional Lie-Poisson system (\ref{eq:LiePoissonContinuous}) to a finite-dimensional Lie-Poisson system was first employed in \cite{zeitlin2004self}. This was motivated by a projection method on a flat geometry in \cite{zeitlin1991finite} for the 2D Euler equations. Since the potential vorticity evolves according to the Lie-Poisson bracket, it is an element of the associated Poisson algebra. The projection method is based on a finite-dimensional approximation of the Lie-Poisson bracket. Since a finite-dimensional Lie algebra can always be written as a matrix algebra, the method consists of constructing a matrix Lie algebra of dimension $(N\times N)$, such that the sequence approximates the infinite-dimensional Poisson algebra as $N$ goes to infinity. This convergence is defined in terms of the structure constants of a basis of the algebra. In other words, a sequence of Lie algebras is said to converge to the infinite-dimensional Poisson algebra $\mathcal{C}^\infty(\mathcal{S}^2)$ if the structure constants of the matrix bases converge to that of the spherical harmonics basis $Y_{lm}$ of $C^\infty(\mathcal{S}^2)$. As was shown for a general case in \cite{bordemann1991gl}, and later for the spherical geometry by \cite{zeitlin2004self} based on the work of \cite{hoppe1989diffeomorphism}, such a sequence exists, given by the basis $T_{lm}$ for the matrix Lie algebra $\mathfrak{u}(N)$, i.e., the algebra of skew-Hermitian matrices. Given a function $a\in C^\infty(\mathcal{S}^2)$ written in terms of the basis as $a=\sum_{l=0}^\infty \sum_{m=-l}^l a_{lm} Y_{lm}$, the implied projection $\Pi_N: \mathcal{C}^\infty(\mathcal{S}^2)\to \mathfrak{u}(N)$ is then given by

\begin{equation}\label{eq:projection}
    \Pi_N a = A:= \sum_{l=0}^{N-1} \sum_{m=-l}^l i a_{lm} \Hat{T}_{lm},
\end{equation}

where the modified basis $\hat{T}_{lm}$ follows directly from the complex part of $T_{lm}$ as detailed in \cite{cifani2023efficient}. Similar to their infinite-dimensional counterparts, $\hat{T}_{lm}$ are eigenfunctions of the Laplace operator, which, in this case, is the discrete Laplace operator on the sphere $\Delta_N$ as provided by \cite{hoppe1998some}. Thus, the basis matrices are calculated by solving the eigenvalue problem

\begin{equation}\label{eq:basismatrices}
    \Delta_N \hat{T}_{lm} = l(l+1) \hat{T}_{lm}
\end{equation}

Although the discrete Laplacian is a fourth-order tensor, it operates independently on diagonals of a matrix. Thus, the tensor splits into $(2N-1)$ blocks $\Delta_N^{(m)}$ of size $N-|m|$. Moreover, these blocks are tridiagonal given as

\begin{equation}\label{eq:LaplacianTriBlock}
\begin{aligned}
    (\Delta_N^{(m)})_{ij} =&\, 2 \delta_i^j \left( s(2i+1+m) - i(i+m) \right)  \\&-(\delta_{i+1}^j + \delta_{i}^{j+1}) \sqrt{(i+1+m)(i+1)}\sqrt{(N-1-i-m)(N-1-i)},
\end{aligned}
\end{equation}
where $s=(N-1)/2$ and $i,j = 0,\ldots, N-m-1$. This sparse structure of the discrete Laplace operator can be inverted using $\mathcal{O}(N^2)$ operations, which has enabled efficient numerical solvers such as reported in \cite{cifani2023efficient} on which the current numerical framework is built. As the discrete Laplacian splits into independent blocks acting along diagonals of a matrix, it follows that the matrices $T_{lm}$ can be represented as $(N-|m|)$-dimensional complex vectors, occupying the $m$-th diagonal of an otherwise empty matrix.

The similarity between the structure constants of the discrete basis $T_{lm}$ and those of $Y_{lm}$ in the continuous case also results in a close relation between the Poisson bracket (\ref{eq:PoissonBracketDefinition}) and the Lie bracket on $\mathfrak{u}(N)$, which is the matrix commutator. The relation is given by
\begin{equation}
    \Pi_N \{f,g\} = [\Pi_N f, \Pi_N g]_N + \mathcal{O}(N^{-2}),
\end{equation}
where $[A,B]_N= N^{3/2}/\sqrt{16\pi}\,(AB-BA)$ is the rescaled matrix commutator (\cite{bordemann1991gl,bordemann1994toeplitz}. Using the explicit projection in (\ref{eq:projection}), we introduce the layerwise potential vorticity matrices $Q_j$ and associated stream matrices $P_j$ using
\begin{equation}
    Q_j = \sum_{l=0}^{N-1} \sum_{m=-l}^l i(q_j)_{lm} \hat{T}^{(N)}_{lm}, \quad \mbox{and} \quad P_j = \sum_{l=0}^{N-1} \sum_{m=-l}^l i (\psi_j)_{lm} \hat{T}^{(N)}_{lm}, \quad j=1,\ldots, M
\end{equation}
Similarly, we introduce the Coriolis matrix $F:=\Pi_N f$ and its square $S:=\Pi_N f^2$, which allows us to derive the spatial discretization of (\ref{eq:PVtransportPoisson}) and (\ref{eq:multi-layerPsi}) as
\begin{equation}\label{eq:PVevolutionQuantized}
    \dot{Q}_j = -[P_j, Q_j] \quad j = 1,\ldots,M
\end{equation}
with potential vorticity given in terms of the stream matrices as
\begin{equation}\label{eq:PVdefinitionQuantized}
    Q_j = \Delta_N P_j + F + \Tilde{S} \circ \left(\sum_{k=1}^M A_{jk}  P_k\right)\quad j = 1,\ldots,M
\end{equation}
where the projection of the product $f^2\psi$ is given by the operation $\tilde{S}\circ P$ with $\circ$ denoting element-wise multiplication, as was derived in \cite{franken2024zeitlin}. The matrix $\tilde{S}$ is given by
\begin{equation}
    \tilde{S}_{ij} = -\frac{i}{2}\sqrt{\frac{N}{4\pi}}(S_{ii}+S_{jj}).
\end{equation}
System (\ref{eq:PVevolutionQuantized}) constitutes $M$ isospectral equations. They are Lie-Poisson with respect to the Lie algebra $\mathfrak{u}(N)$ with the Lie bracket $[\cdot,\cdot]$, for which the Casimirs $C_{j,k}$ by isospectrality are given as
\begin{equation}\label{eq:CasimirsDiscrete}
    C_{j,k} = \mbox{trace} \left((Q_j)^k\right), \quad k=0,1,\ldots,N-1, \quad j = 1,\ldots, M,
\end{equation}
which indicates that integrated monomial functions of potential vorticity are conserved in each layer, similar to the dynamics of a single-layer QG model. 

Solving the system of equations (\ref{eq:PVevolutionQuantized}) now requires a suitable time integration method, as well as a method for calculating the stream matrices $P_j$ from the implicit relation given in (\ref{eq:PVdefinitionQuantized}), which is discussed in the following two subsections.

\subsection{Calculation of the streamfunctions}

Since the main dynamical variable in the evolution of multi-layer quasi-geostrophic flow is the potential vorticity, the resulting velocity field, which in turn is calculated from the stream function, must be determined using the implicit relation (\ref{eq:PVdefinitionQuantized}). In the special case where the interaction matrix $A$ is zero, i.e., in the absence of density stratification, the horizontal velocity field in each layer is independent and the stream matrix can be determined using
\begin{equation}\label{eq:PcalculationSingleLayer}
    P_j = (\Delta_N)^{-1}(Q_j-F), \quad j = 1,\ldots, L.
\end{equation}
Similar to the observation that the streamfunction is uniquely determined only up to a constant function on the sphere, the discrete Laplacian is only invertible when adding an additional constraint on the stream matrix. For convenience, we may set $\int_{S^2}\psi_j=0$, $\forall j=1,\ldots,M$, which corresponds in the discrete case to the constraint
\begin{equation}
    \mbox{trace} (P_j) = 0, \quad j=1,\ldots,M.
\end{equation}
Using the diagonal splitting of the discrete Laplacian, the calculation (\ref{eq:PcalculationSingleLayer}) can be performed using $\mathcal{O}(N^2)$ operations following \cite{cifani2023efficient}.

Under the influence of density stratification, the stream matrices no longer decouple between layers, and the fully coupled system (\ref{eq:PVdefinitionQuantized}) needs to be considered. For a general interaction matrix, this involves solving a system of $M$ coupled elliptic equations of size $N\times N$. However, in many cases, the matrix $A$ is diagonalizable, leading to an ideal pre-conditioner for the coupled system.

If the matrix $A$ is diagonalizable, this means that there exists an invertible square matrix $V$ and a diagonal matrix $D$ such that
\begin{equation}\label{eq:modedecomposition}
    A = V D V^{-1}.
\end{equation}
Substitution of this decomposition into equation (\ref{eq:PVdefinitionQuantized}) then gives
\begin{equation}\label{eq:PVmodes}
    \hat{Q}_k = \Delta_N \hat{P}_k + F + \Tilde{S} \circ \left( D_{kk} \hat{P}_k \right), \quad k=1,\ldots, M,
\end{equation}
where $\hat{Q}_k = \sum_{j=1}^M V^{-1}_{kj} Q_j$ and $\hat{P}_k = \sum_{j=1}^M V^{-1}_{kj} P_j$ are the potential vorticity matrices and stream matrices respectively corresponding to the modes of the system~\cite{thiry2024mqgeometry}. Crucially, the potential vorticity matrix $\hat{Q}_j$ associated with a particular mode $j$ is independent of the other modes. Thus, system (\ref{eq:PVmodes}) constitute $M$ independent implicit equations for the stream matrices $\hat{P}_j$, which can be made explicit by solving
\begin{equation}\label{eq:Pmodesolve}
    \hat{P}_k = \left( \Delta_N + D_{kk}\Tilde{S} \circ \right)^{-1} (\hat{Q}_k-F), \quad k = 1, \ldots, M,
\end{equation}
which can be solved efficiently in just $\mathcal{O}(N^2)$ operations using the scheme based on diagonal splitting as developed in~\cite{franken2024zeitlin}. Thus, the preconditioner is ideal in the sense that the system can be solved as efficiently as in the decoupled case. The layerwise PV and streamfunction matrices can then be reconstructed using the relations
\begin{equation}\label{eq:modetolayer}
    Q_j = \sum_{k=1}^M V_{jk}\hat{Q}_k, \quad P_k = \sum_{k=1}^M V_{jk}\hat{P}_k
\end{equation}
In the case where the interaction matrix $A$ is not diagonalizable, there is no such decomposition into independent modes. However, the stream matrices can still be calculated directly from equation (\ref{eq:PVdefinitionQuantized}) using diagonal splitting. If we denote the $m$-th diagonal of a matrix by the superscript $\cdot^{(m)}$ as a row vector, we may introduce the column vectors $P^{(m)},Q^{(m)}\in\mathds{C}^{(N-|m|)L}$ as
\begin{equation}
    P^{(m)} = \begin{bmatrix} P_1^{(m)} & P_2^{(m)} & \cdots & P_L^{(m)} \end{bmatrix}^\top
\end{equation}
and
\begin{equation}
    \mathds{Q}^{(m)} = \begin{bmatrix} Q_1^{(m)} & Q_2^{(m)} & \cdots & Q_L^{(m)} \end{bmatrix}^\top, 
\end{equation}
which allows us to decompose the entire system (\ref{eq:PVdefinitionQuantized}) into its diagonals as
\begin{equation}
    (\mathds{L}^{(m)} + \mathds{S}^{(m)})P^{(m)} = Q^{(m)}-F^{(m)}, \quad  m = -(N-1),\ldots,N-1
\end{equation}
using the matrix operators $\mathds{L}^{m},\mathds{S}^{m}\in \mathds{C}^{[(N-|m|)M]\times[(N-|m|)M]}$ given as 
\begin{equation}
    \mathds{L}^{(m)} = I_M \odot \Delta_N^{(m)}
\end{equation}
and
\begin{equation}
    \mathds{S}^{(m)} = A \odot \Tilde{S}^{(m)}
\end{equation}
where $\odot$ denotes the Kronecker product and $I_M$ is the identity matrix of dimension $M$. The matrix $(\mathds{L}^{(m)} + \mathds{S}^{(m)})$ is extremely sparse with only 5 nonzero diagonals as can be seen from equations (\ref{eq:interactionmatrix}) and (\ref{eq:LaplacianTriBlock}). The system can be solved efficiently using appropriate pre-conditioners~\cite{chen2005matrix} or using iterative methods. Here we note that in the case of a diagonalizable matrix $A$, the preconditioner can be found from the decomposition (\ref{eq:modedecomposition}), which leads directly to system (\ref{eq:Pmodesolve}). Note that one only needs to solve for the diagonals $m\geq 0$, since the matrices $P_j$ are skew-Hermitian. The linear scaling of the computational complexity with the number of layers facilitates simulations at high horizontal resolutions, similar to previous work on barotropic QG flow as reported by \cite{franken2024zeitlin}.

\subsection{Time integration method}

The evolution of the dynamical system is carried out by solving equation (\ref{eq:PVevolutionQuantized}) in time. Analogous to the conservation of integrated functions of potential vorticity in the continuous system, this system of ODEs preserves the trace of polynomial functions of $Q_j$ up to polynomial order $(N-1)$ as indicated in equation (\ref{eq:CasimirsDiscrete}) for each layer. This property is equivalent to the spectrum of the layer-wise potential vorticity matrix being conserved~\cite{watkins1984isospectral}, indicating that the evolution in (\ref{eq:PVevolutionQuantized}) is iso-spectral.

This property is similar to the isospectral evolution of the potential vorticity matrix in a single-layer QG model as in~\cite{franken2024zeitlin}, where it was also remarked that the isospectral property holds independent of the choice of advecting stream matrix. Therefore, we can select the same time-integration method to conserve the matrix spectrum on a discrete level.

We employ the time integrator as formulated in \cite{modin2020casimir}, based on second-order symplectic midpoint integration. This integrator falls within the class IsoSyRK methods, which are both symplectic and isospectral~\cite{modin2020lie}. Given the solution $Q_j^n$ at time $t_n$, the solution of (\ref{eq:PVevolutionQuantized}) at time $t_{n+1} = t_n + h$ can then be written as follows

\begin{equation}\label{eq:TimeIntegrator}
\begin{cases}
    \Tilde{Q}_j = Q_j^n + \frac{h}{2}\left[P_j,\Tilde{Q}_j^n\right] + \frac{h^2}{4} P_j \Tilde{Q}_j P \\
    Q_j^{n+1} = \Tilde{Q} + \frac{h}{2}\left[ \Tilde{P}_j, \Tilde{Q}_j \right] - \frac{h^2}{4}\Tilde{P}_j \Tilde{Q}_j \Tilde{P}_j,
\end{cases}
\end{equation}

where $\Tilde{P}_j$ is the stream matrix corresponding to the potential vorticity matrix $\Tilde{Q}_j$ as given by equation (\ref{eq:PVdefinitionQuantized}). The integration thus consists of two steps. The first equation gives an implicit relation for the intermediate step $\Tilde{Q}_j$, which is solved using fixed point iteration. This creates a sequence $\left\{ \Tilde{Q}_j^k \right\}$ with $\Tilde{Q}_j^0 = Q_j^n$. The time step $h$ is chosen sufficiently small such that the sequence converges to $\Tilde{Q}_j$ as $k\to \infty$. This contraction can be performed in parallel for each layer, with each process terminating when the sequence has converged to within a set tolerance. For the chosen time step sizes $h$, the $L_\infty$ norm of $\Tilde{Q}_j^{k+1}-\Tilde{Q}_j^k$ typically reaches $\mathcal{O}(10^{-8})$ after only 4 or 5 iterations. After sufficient convergence, the stream matrices $\Tilde{P}_j$ are calculated, and a full time step is completed by evaluating the right-hand side in the second equation in (\ref{eq:TimeIntegrator}).

\section{Simulation of multi-layer geostrophic turbulence}\label{sec:Simulations}

In this section, we demonstrate the capabilities of the developed numerical method by simulating two types of geophysical flows. In Section \ref{subsec:zonaljets}, we study the emergence of zonally elongated structures in the flow from random initial data in an energy- and entropy-conserving setting for a six-layer model. Section \ref{subsec:forcedturb} presents results of forced geostrophic turbulence for a three-layered setting, where we compare the resulting velocity fields and energy spectra to those of single-layer QG turbulence. 

\subsection{Zonal jet formation}\label{subsec:zonaljets}

To verify the structure-preserving properties of the numerical methods, we perform a simulation of geostrophic turbulence in the absence of forcing, dissipation and friction. We consider an aqua planet (an idealized planet without topographic features~\cite{blackburn2013context}) of radius $R=1000$ km, rotating with a period of $10^4$ seconds. For a typical rms velocity of $0.2$ m/s achieved in this simulation, this leads to a global Rossby number of $\mbox{Ro}=V_{rms}/R\Omega \approx 3\cdot 10^{-4}$, which is well within the geostrophic regime~\cite{luesink2024geometric}. For illustration purposes, the fluid layer is decomposed into 6 layers of equal thickness set to $H_j = 2$ km. A strong stratification is imposed between the layers by setting the reduced gravity at the layer interfaces to $0.8, 0.6, 0.4, 0.2$ and $0.1$ from top to bottom, indicating that the density gradient is largest in the top layers. This choice of parameters leads to baroclinic Rossby radii of $91, 45, 32, 24$ and $15$ km respectively.

A random initial large-scale velocity field is imposed in the top layer, and similarly at smaller amplitudes in lower layers. Following the work of \cite{cifani2023efficient}, this is done by selecting the spherical harmonic modes with $1<l<30$ of all stream functions and giving them a random phase and a normally distributed amplitude around the value $\psi_0/l(l+1)$. The reference value $\psi_0=2\cdot 10^{-4}/j$ is tuned such that the maximum jet velocity in the top layer is close to unity, with $j$ the layer number ensuring that lower layers have a lower starting velocity scale. With this setup, the radial (vertical) kinetic energy transfer can be studied as the flow develops and kinetic energy is exchanged between the different layers. We simulate at a horizontal resolution of $N=128$ during a period of $3\cdot 10^4$ 'days' (number of revolutions of the sphere), with an integration time step of $\Delta t = 10^3$ s, which implies a temporal resolution of $10$ time steps per 'day'.

\begin{figure}
    \centering
    \includegraphics[width=\columnwidth]{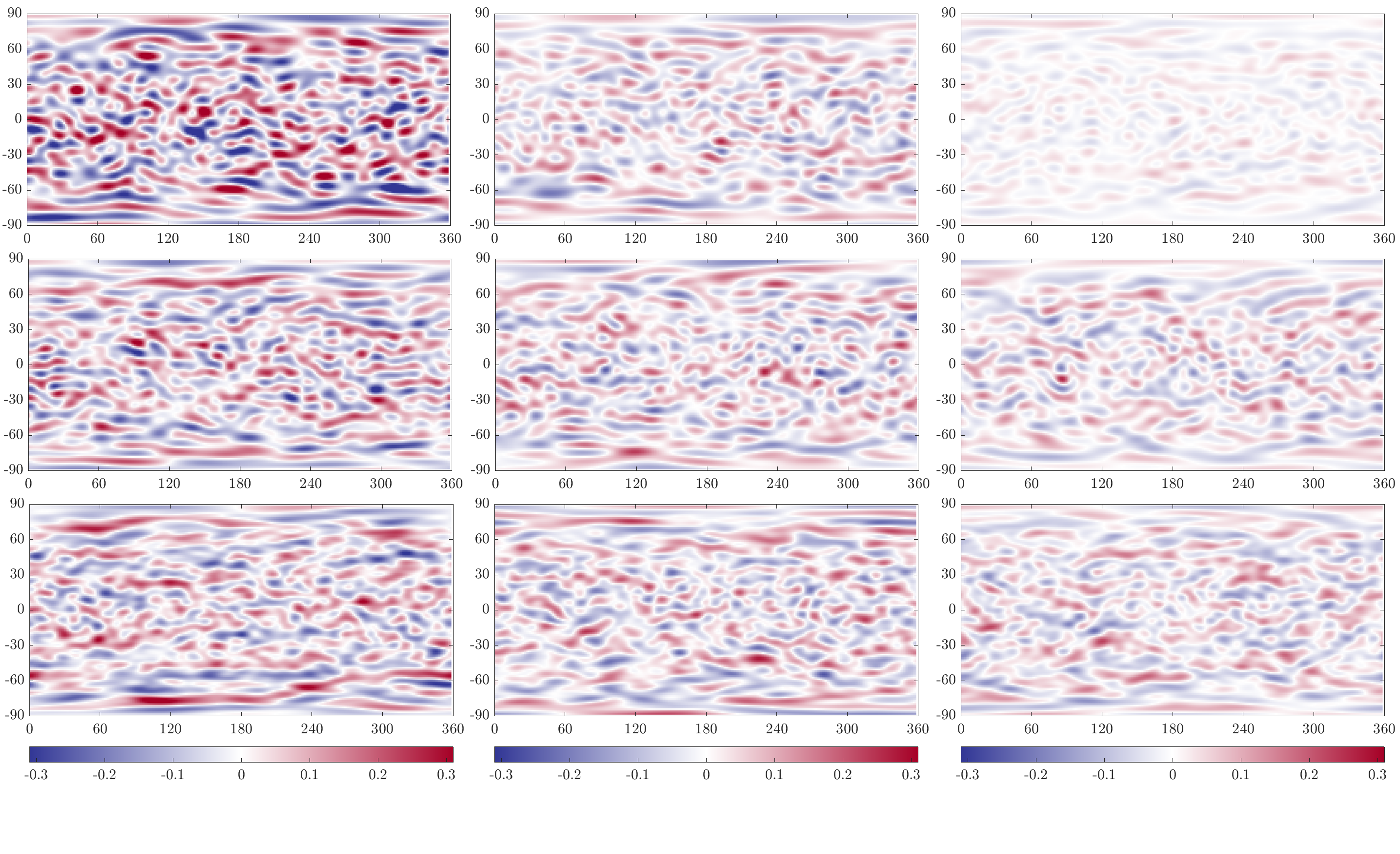}
    \caption{Instantaneous zonal velocity fields for the energy and enstrophy-conserving setup. From left to right, the velocity fields of layers 1, 2 and 6 are visualized on a latitude-longitude grid. The velocity fields are calculated at $t=0$ (top), $t=10$ days (middle) and $t=200$ days (bottom).}
    \label{fig:Hyperbolic_table}
\end{figure}

Figure~\ref{fig:Hyperbolic_table} shows several snapshots of the zonal component of the velocity field. The top row shows the initial zonal velocity of layers 1, 2 and 6, which are the top layer, second layer and bottom layer respectively. These initial velocity fields contain all spherical harmonics modes with degree $l\leq 30$, with a decreasing amplitude for lower layers. As the fluid evolves, two effects are visible in the zonal velocity snapshots. The second row in Figure~\ref{fig:Hyperbolic_table} shows the snapshots after 2500 days, while the third row shows a snapshot after 30 000 days. The first effect is a migration of kinetic energy from the top layer to lower layers due to the interaction between layers. Layerwise, this represents a shift in the energy level at a constant Casimir surface. In other words, kinetic energy is transported between layers whilst enstrophy is conserved in each layer. The second effect is the evolution of potential vorticity within each layer. This leads to an elongation of vortex structures in the zonal direction, which is particularly visible in the equatorial region. Furthermore, due to the nonlinear interactions between large-scale modes, small-scale structures are formed with $l>30$. Notably, due to the strong imposed stratification, the elongation of vortices in the east-west direction continues toward the formation of zonal jets in this particular case. In this simulation, the baroclinic Rossby radii are large compared to the radius of the sphere such that baroclinic effects dominate the flow. The influence of baroclinic effects on jet formation is not well understood (see for example \cite{lee2005baroclinic,gilet2009nonlinear}). Further study is of great interest and suggested for future research.

\begin{figure}
    \centering
    \includegraphics[width=.6\columnwidth]{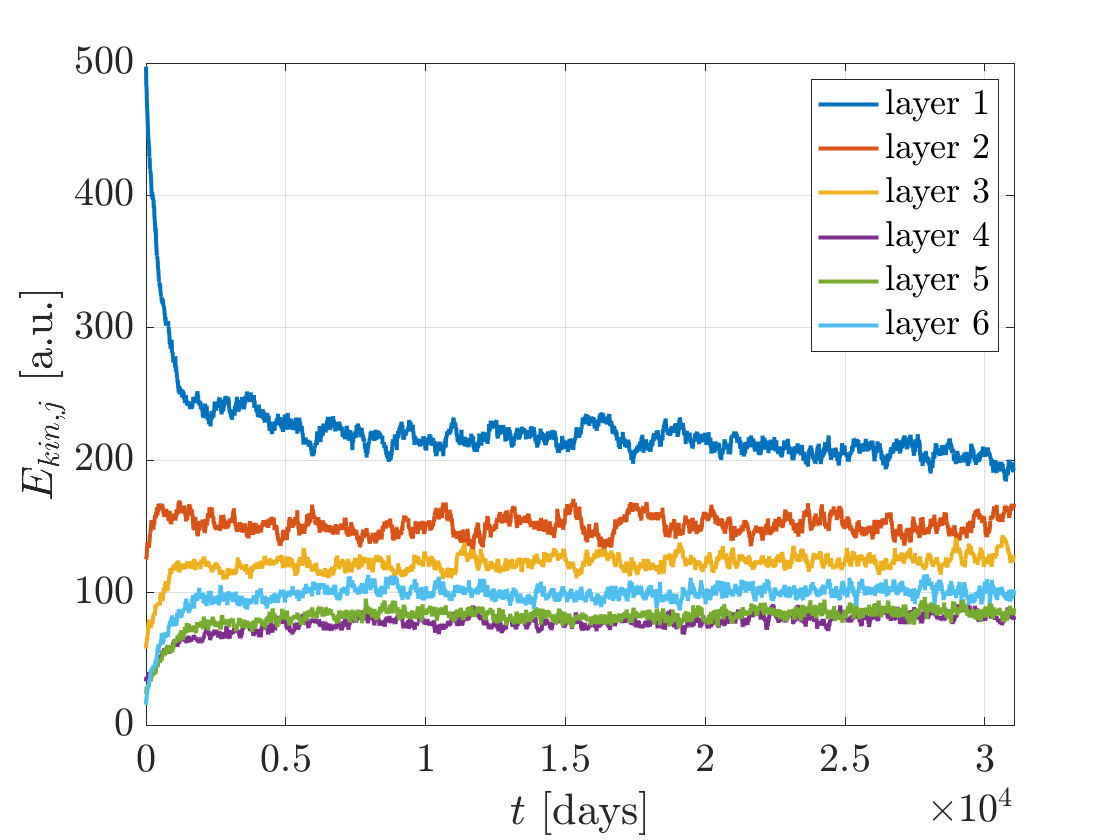}
    \caption{Time evolution of the layer-wise kinetic energy in arbitrary units. The initial condition is chosen such that $E_{kin,j}\sim j^{-2}$, where $j$ is the layer number. As the system evolves, kinetic energy is redistributed between the layers due to the interaction at the interfaces, and after approximately 250 days, the flow settles into a statistically stationary state with a stable kinetic energy distribution.}
    \label{fig:Hyperbolic_Ekin_table}
\end{figure}

The transfer of kinetic energy between layers is visualized in Figure~\ref{fig:Hyperbolic_Ekin_table}, showing the kinetic energy in each layer as a function of time. The initial total energy per layer is determined via the amplitude of the random initial fields, leading to an initial kinetic energy proportional to $j^{-2}$ where $j$ is the layer number. The figure shows a transient phase up to approximately 250 days of simulation during which kinetic energy is gradually transferred from the top layers towards the lower layers. After that, a stable distribution was reached which persists even after a very long time integration in which the top layers are only marginally more dynamically active than the lower layers due to the strong interaction between layers.

\begin{figure}
    \centering
    \includegraphics[width=\columnwidth]{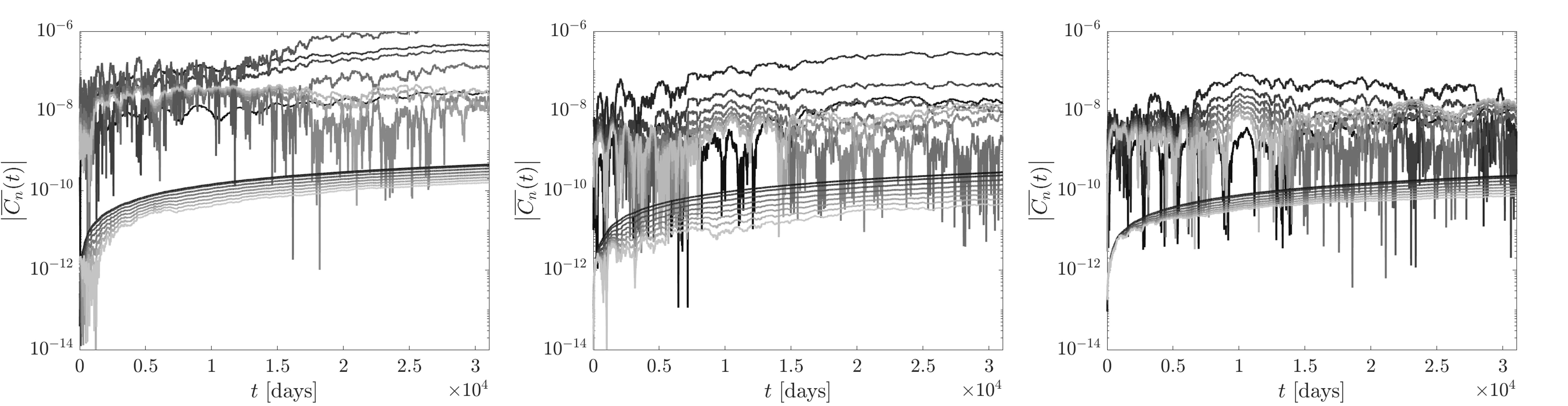}
    \caption{Evolution of the relative error in the first 16 Casimirs in the system for the top layer (left), second layer (middle) and bottom layer (right). Successively higher order Casimirs are plotted using a lighter hue of gray, with even-order Casimir clustering around a relative error of $10^{-10}$ and odd-order Casimir displaying a relative error on the order of $10^{-8}$.}
    \label{fig:Hyperbolic_Casimirs_table}
\end{figure}

Finally, we illustrate the Casimir conserving property of the numerical method in Figure~\ref{fig:Hyperbolic_Casimirs_table}, which shows the time evolution of several Casimirs for layers 1, 2 and 6 (from left to right). The figures show the relative error in the value of the first 16 Casimirs as a function of time, as given in equation (\ref{eq:CasimirsDiscrete}). The relative error of the Casimirs is given in terms of their respective initial values and is displayed using increasingly light hues of grey for higher-order Casimirs. We find similar results as those reported in \cite{franken2024zeitlin} for single-layer QG dynamics in the Casimir-preserving setting. The even-order monomial Casimirs are conserved up to machine precision in each, which is close to $10^{-10}$ in this case. This includes the second-order Casimirs, representing the total enstrophy in each layer. The odd-order Casimirs display an error of approximately $10^{-8}$. This is of a similar magnitude as the tolerance of the fixed point iteration used in the time integrator (\ref{eq:TimeIntegrator}), as well as the numerical accuracy achieved in the procedures for generating the basis matrices $T_{lm}^{(N)}$ in equation (\ref{eq:basismatrices}). These results thus confirm that the numerical method conserves all monomial Casimirs of the discrete system up to a very high precision.

\subsection{Forced turbulence}\label{subsec:forcedturb}

Many geophysical flows, such as the Earth's ocean and atmosphere, are heavily dependent on external forcing and are subject to dissipative effects in potential vorticity~\cite{chai2016understanding}. Examples of external forcing are wind shear on the upper ocean or thermal forcing from radiation~\cite{vallis2017atmospheric}. Dissipation of potential vorticity occurs beyond the geostrophic regime due to unresolved physical processes at small length scales. This effect is typically modelled using an eddy viscosity term, see for example~\cite{shevchenko2015multi,zalesny2022variational,carigi2023dissipation}. Furthermore, in the case of modelling oceanic flow or the wind layer on other planets, the interaction between the bottom layer and the unresolved deep ocean is modelled with an Ekman boundary layer which corresponds to a linear drag force on the bottom layer~\cite{zeitlin2018geophysical,li2023stochastic}. These terms can be introduced into the PV evolution equation (\ref{eq:PVtransportPoisson}) as follows~\cite{shevchenko2015multi}

\begin{equation}\label{eq:PVevolutionwithdrag}
    \dot{q}_j + \left\{ \psi_j, q_j \right\} = \delta_{1,j} F_w - \delta_{j,N} \mu \Delta \psi_j + \nu \Delta^2 \psi_j,
\end{equation}

where $\delta_{i,j}$ is the Kronecker symbol, $F_w$ is the (wind) forcing acting directly on the top layer only, $\mu$ is the Eckman drag coefficient for the bottom friction layer and $\nu$ is the eddy viscosity. Similar to the approach taken in~\cite{thiry2024mqgeometry}, we integrate the terms on the right-hand side of (\ref{eq:PVevolutionwithdrag}) using a third-order total variation diminishing Runge-Kutta (TVD-RK3) scheme. Given an ordinary differential equation
\begin{equation}
    \dot{y}(t) = f(y(t)),
\end{equation}
the scheme can be written as the following three-stage process
\begin{equation}
    \begin{split}
        y^{(1)} &= y^{(0)} + h f(y^{(0)})\\
        y^{(2)} &= y^{(1)} + \frac{h}{4} \left( f(y^{(1)}) - 3f(y^{(0)}) \right) \\
        y^{(3)} &= y^{(1)} + \frac{h}{12} \left( 8f(y^{(2)}) - f(y^{(1)}) - f(y^{(0)})\right),
    \end{split}
\end{equation}
where $h$ is the adopted time step size. The full system (\ref{eq:PVevolutionwithdrag}) can then be solved using an operator splitting method. By denoting the isospectral time integration map (\ref{eq:TimeIntegrator}) for the advection term as $\varphi_{iso,h}$ for a time step of size $h$, and similarly denoting the TVD-RK3 integration map with $\varphi_{RK,h}$, we use a second-order Strang operator splitting to complete a full time stepping map $\varphi_h$ using the composition

\begin{equation}\label{eq:StrangSplitting}
    \mathbf{Q}^{k+1} = \left( \varphi_{RK,h/2} \circ \varphi_{iso,h} \circ \varphi_{RK,h/2} \right) \mathbf{Q}^k
\end{equation}

where $\mathbf{Q}^{k}$ denotes the solution at time $t_k$, where $t_{k+1} = t_k + h$. Using this composition, the full dynamical equation (\ref{eq:PVevolutionwithdrag}) can be solved while retaining the structure-preserving integration for the nonlinear advection term, even though the full system does not conserve energy or Casimirs due to the physical forcing and dissipation processes.

For this numerical experiment, we localize the forcing $F_w$ around a narrow band in spectral space around harmonics of degree $l=50$ as is often used for studying homogeneous turbulence~\cite{boffetta2010evidence,cifani2022casimir}. The remaining parameters of this simulation are based on the double-gyre experiment reported in \cite{thiry2024mqgeometry}. That experiment aims to simulate the emergence of a double wind-driven gyre on a tangent plane to study the westward jet between the gyres emanating from the eastern domain boundary. In our global fluid domain, the absence of north-south boundaries prevents replicating this experiment in all its detail. Therefore, we use the aforementioned isotropic forcing to induce turbulent flow in the top layer, and the results for the multilayer system will be compared to the single-layer QG model presented in \cite{franken2024zeitlin}. 

The fluid domain is considered to represent the Earth, which leads to $R=6\cdot 10^6$ m and a rotational period of the planet of $T = 86\,400$ s. The experiment models oceanic flow with typical horizontal velocities of $0.5$ m/s, leading to a global Rossby number of $1\cdot 10^{-3}$. We consider a three-layer system with a layer thickness of $H_1=400$ m, $H_2 = 2000$ m and $H_3 = 4000$ m. The reduced gravity of the interfaces are set to 0.4 and 0.2 m/s$^2$, leading to baroclinic Rossby radii of $152$ and $249$ km respectively. The eddy viscosity is set to $\nu=60$ m$^2$/s following \cite{cifani2022casimir}, and the Ekman drag coefficient is set to $\mu = 0.01$ s$^{-1}$ similar to \cite{thiry2024mqgeometry}. The spatial resolution is set to $N=1024$ such that the smallest resolved length scale is approximately $10$ km. This is significantly smaller than the baroclinic Rossby radii, thus ensuring that baroclinic waves are well-resolved~\cite{shevchenko2015multi}. The time step size is set to $10^3$ seconds, resulting in approximately 87 time steps per day, with a total simulation time set to 2100 days. For comparison purposes, we also simulate a single-layer QG system with the same parameters by setting the number of layers to 1. The influence of the geostrophic balance is maintained by retaining the value of $A(1,1)$ in equation (\ref{eq:multi-layerPsi}). This is equivalent to a one-way interaction of the (single) top layer with a stationary deep ocean.

\begin{figure}
    \centering
    \includegraphics[width=\columnwidth]{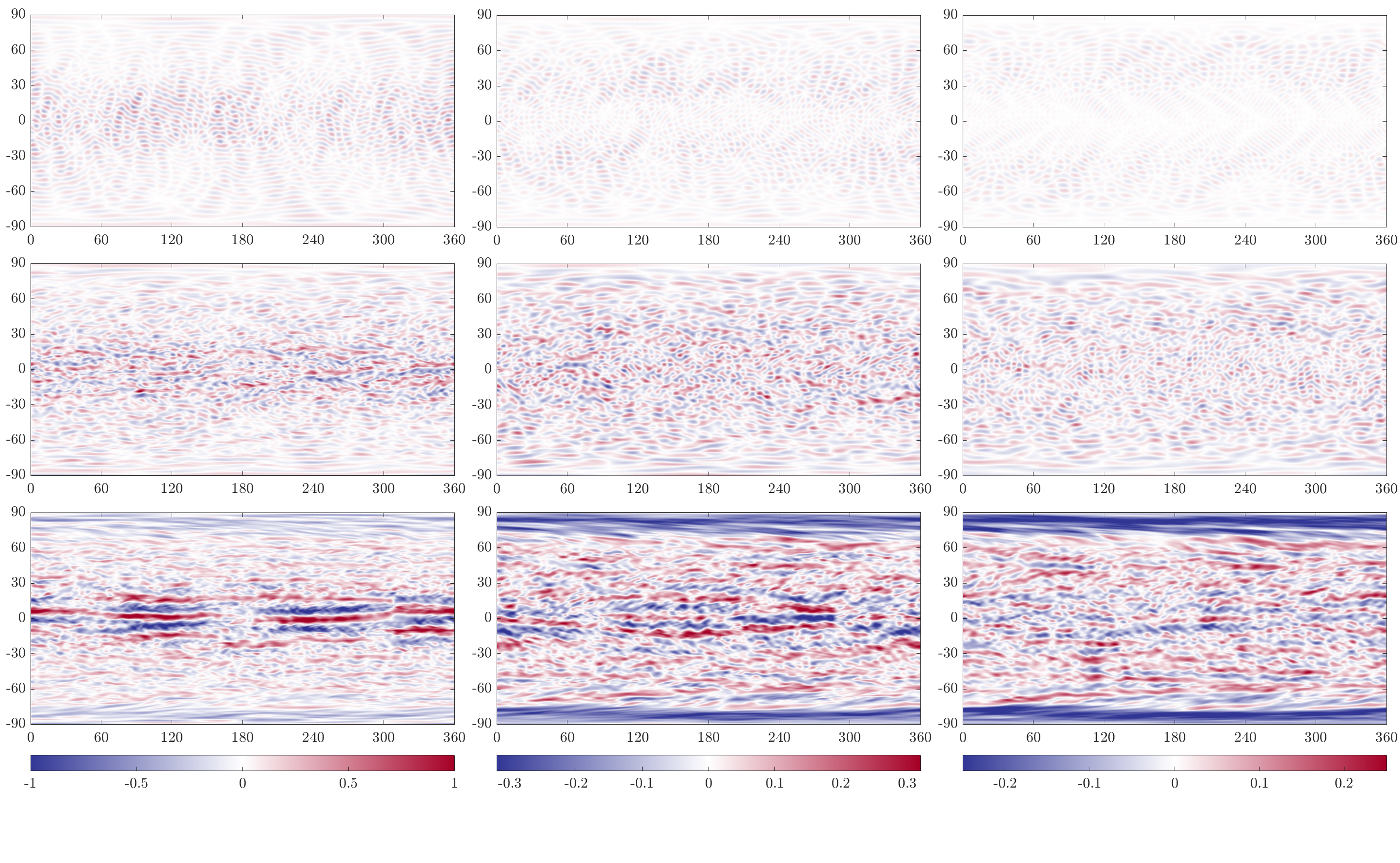}
    \caption{Snapshots of the zonal velocity field. From top to bottom, snapshots are taken at $T=140$ days, $T=700$ days and $T=2100$ days. From left to right, snapshots of the top layer, middle layer and bottom layer are shown respectively. Red colour indicates prograde motion, while blue colour represents retrograde motion, with a distinct colour scale for each layer to improve visibility.}
    \label{fig:ForcedDissipative_table}
\end{figure}

Figure~\ref{fig:ForcedDissipative_table} shows snapshots of the zonal component of the layer-wise velocity fields at several times. The flow develops from an initially stationary situation due to the random forcing on the top layer. In the figure, the left column shows the top layer zonal velocity field, the middle column shows the second layer and the right column shows the bottom layer. Note the different colour scales for the columns to visualize the resulting flow fields more clearly. The top row shows a snapshot after 140 days, which shows the development of small-scale flow structures at the length scale of the random forcing. The middle row shows the solution after 700 days, at which point the top layer clearly shows the emergence of elongated structures in the zonal direction. These develop further into the clear zonal jets shown after 2100 days on the bottom row of the figure. 

As is clear from the simulations, the top layer is the most dynamically active as it is directly influenced by the external forcing and has the lowest inertia due to the smaller layer thickness. It clearly shows the emergence of zonal jets in the equatorial region, which is a well-known feature of geostrophic flows~\cite{scott2012structure}. Away from the equator, no zonal jets form and the flow is qualitatively isotropic. The lower two layers are indirectly forced by the upper layer. This forcing appears as a PV source when the streamfunctions of two layers differ. Since the interaction term as seen in (\ref{eq:multi-layerPsi}) is proportional to $f^2$, this source term is always smallest near the equator, which explains the low velocities in these regions after short simulation times as seen in the top row of Figure~\ref{fig:ForcedDissipative_table}. As the flow develops further, the internal dynamics of the lower layers result in a largely homogeneous flow field. 

\begin{figure}
    \centering
    \includegraphics[width=.6\columnwidth]{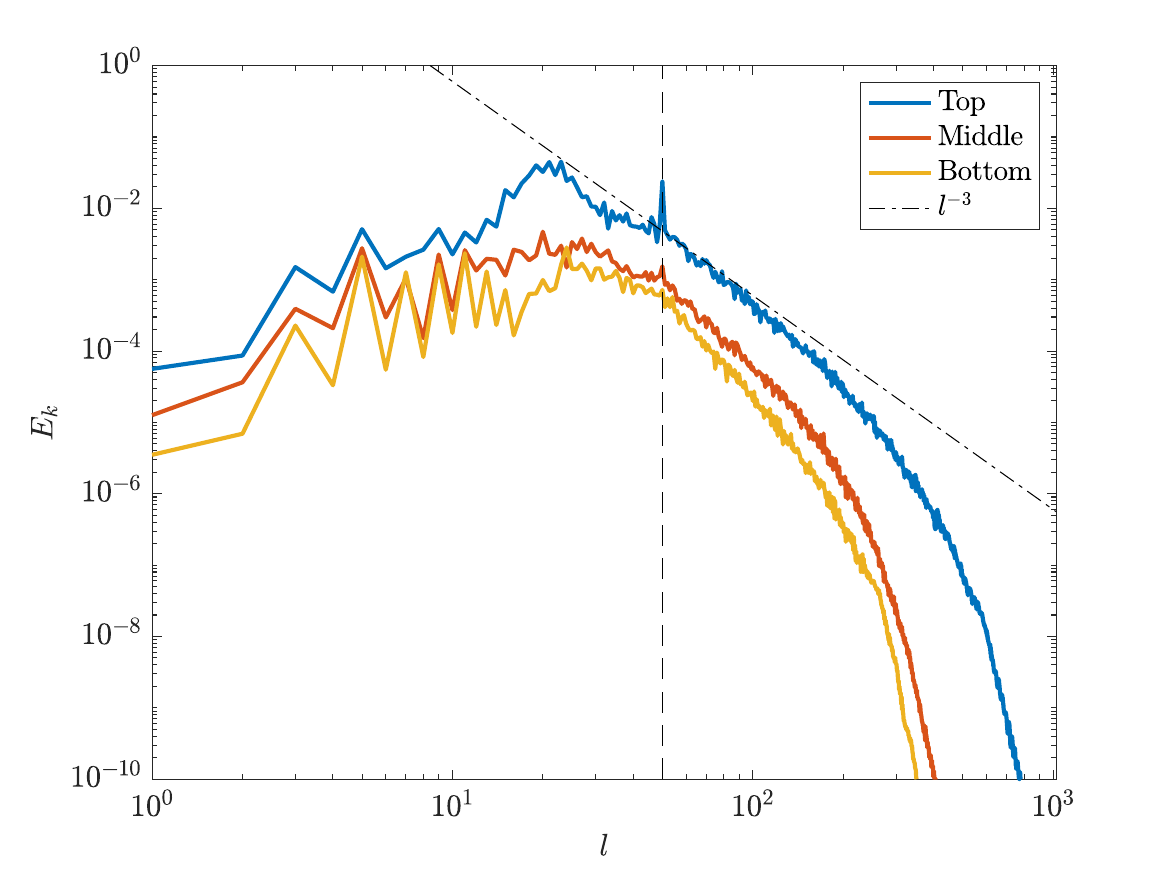}
    \caption{Kinetic energy profiles for all three layers at $T=2100$ days. For reference, the forcing wavenumber $l=50$ is denoted with a dashed line, and the characteristic slope of $l^{-3}$ is shown for reference.}
    \label{fig:Ekprofiles}
\end{figure}

While zonal jets form the dominant flow feature in the top layer, the lower layers show a more homogeneous distribution of kinetic energy in the flow domain. This is also clearly visible in the kinetic energy spectra of the layer-wise velocity fields shown in Figure \ref{fig:Ekprofiles}. In blue, the kinetic energy spectrum of the top layer is shown. The external forcing at $l=50$ is visible in the spectrum, from which energy is distributed to all other modes. Furthermore, most energy is contained in harmonics with degree $l=10-20$, which corresponds to the scale of the zonal jets. At high wavenumbers, the energy spectrum decays exponentially due to the eddy viscosity dissipation. The kinetic energy spectra of the lower layers exhibit largely the same structure as the top layer. However, the modes corresponding to the zonal jets have a significantly lower amplitude, indicating that zonal jets do not dominate the flow in these layers. In contrast, most kinetic energy in these layers is contained in the modes with degree $l=20-30$, corresponding to structures with a size between 200 and 300 kilometres, closely corresponding to the baroclinic Rossby radii. 
\begin{figure}
    \centering
    \includegraphics[width=\columnwidth]{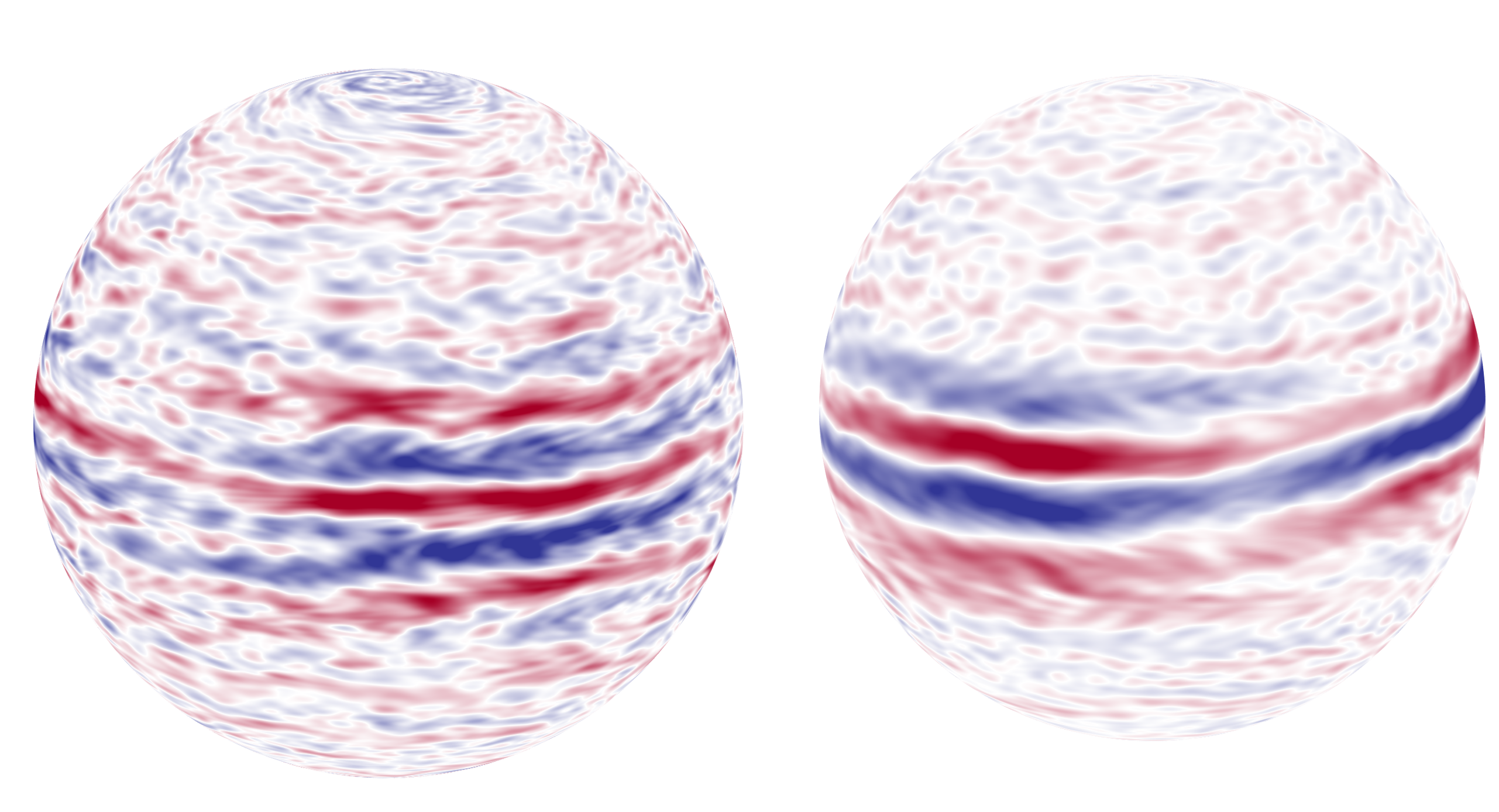}
    \caption{Snapshots of the zonal velocity field. The left picture shows a snapshot of the top layer for the multi-layer simulation shown in figure \ref{fig:ForcedDissipative_table}. The picture on the right shows a snapshot of a fully developed single-layer QG turbulence simulation at the same Rossby deformation radius.}
    \label{fig:ForcedDissipative_comparison}
\end{figure}

In Figure~\ref{fig:ForcedDissipative_comparison}, a snapshot of the top layer zonal velocity field is compared to the simulation of single-layer barotropic QG dynamics with the same parameters as described above. Qualitatively, we find a striking similarity in the emergence of zonal jets. However, the multi-layer simulation shows more complex turbulent structures, hinting at a more dynamically complex interaction between layers. This is in line with observational data from the planet Jupiter, which, besides strong zonal jets in the equatorial region, shows a dynamically active polar region~\cite{adriani2018clusters}. Another striking feature of the multi-layer simulation is the appearance of a stable polar vortex, which is a key feature in many observed planetary atmospheres~\cite{mitchell2021polar}. While these similarities with observational data are only qualitative, the developed numerical framework for simulating stratified geostrophic turbulence enables future quantitative studies on global planetary flows.

\section{Conclusion and outlook}\label{sec:conclusion}

In this paper, we presented a global formulation for the multilayer quasi-geostrophic (QG) equations and developed a Casimir-preserving numerical method for simulations. By extending the baroclinic QG models from traditional local $f$-plane or $\beta$-plane approximations to a global spherical framework, this work provides a comprehensive tool for studying large-scale atmospheric and oceanic dynamics.

The numerical method is based on similar global methods for the barotropic QG equations developed in \cite{franken2024zeitlin}, which in turn extends a study on the two-dimensional Euler equations in \cite{cifani2023efficient}. In the current work, we developed a global formulation of the multilayer QG equation and applied a Casimir-preserving discretization to the resulting two-dimensional transport equations. This led to a finite-dimensional system which preserves all monomial Casimirs of the discrete system. Using an isospectral symplectic time integrator, we show that all discrete Casimirs are preserved up to a relative error of $10^{-8}$.

We have furthermore employed the developed method to simulate forced geostrophic turbulence. We have shown that zonal jets form at the top layer, which agrees with simulations on a single-layer model. Furthermore, we show from the kinetic energy spectra that the dynamics of the lower layers mostly concentrate near the respective baroclinic Rossby radii. The development of this novel numerical framework paves the way for application in large-scale simulations of global atmospheric and ocean dynamics. A particular challenge is the development of benchmark simulations for geostrophic flow on a sphere.

\bibliographystyle{plain}
\bibliography{references}

\end{document}